\magnification= \magstep1
\hsize 6.5 true in
\vsize 8.5 true in
\voffset=.6in
%\baselineskip=14pt
%\nopagenumbers
\pageno=1
\newcount\knum
\knum=2
\newcount\uknum
\uknum=1
\newcount\znum
\znum=1
\newcount\notenumber

\def\Za{\the\znum \global\advance\znum by 1}
\def\ukneu{\znum=1\global\advance\uknum by 1}
\def\ATG{hep-th/9605129}
\def\THA{May 1996}
\headline={\ifnum\pageno<2 \hskip 10.9cm\hbox{\hfill\vbox{
\hbox{\ATG}\hbox{\THA} }}
\else\hfill\fi}

%\headline={\ifodd\pageno\rightheadline \else\leftheadline\fi}
%\def\rightheadline{\vbox{\line{\petit\noindent\hskip -0.7 true cm
%\sl\S 3 Geometrie prinzipaler Einbettungen
%\hfil{\tenrm \folio}}\vskip 0.7cm}}
%\def\leftheadline{\vbox{\line{\petit\noindent\sl\hskip -0.7 true cm
%{\tenrm \folio}\hfil ????}
%\vskip 0.7cm}}
\font\tbfontt=cmbx10 scaled \magstep0
\font\gross=cmbx10 scaled \magstep2
\font\mittel=cmbx10 scaled \magstep1

\font\TT=cmcsc10 scaled \magstep0

\font\eightrm=cmr8
\font\nirm=cmr9
\font\kl=cmr6
\font\eightit=cmti8 scaled \magstep0
\def\sqr#1#2{{\vcenter{\vbox{\hrule height.#2pt\hbox{\vrule width.#2pt
height#1pt \kern#1pt \vrule width.#2pt}\hrule height.#2pt}}}}

\def\Theorem{\vskip 0.3 true cm\sl {\TT Theorem \Za.}{ }}

\def\Corollary{\vskip 0.3 true cm\sl {\TT Corollary \Za.}{ }}
\def\Lemma{\vskip 0.3 true cm\sl {\TT Lemma \Za.}{ }}

\def\cz{{\rm C\hskip-4.8pt\vrule height5.8pt\hskip5.3pt}}
\def\kz{{\rm C\hskip-4.0pt\vrule height4.3pt\hskip5.3pt}}  %Index C-Zeichen
\def\rz{{\rm I\hskip -2pt R}}
\def\nz{{\rm I\hskip -2pt N}}

\def\Fz{{\rm I\hskip -2pt F}}
\def\zz{{\rm Z\hskip -4pt Z}}
  %quaternionisches Zahlenzeichen
\def\eins{{\rm 1\hskip -2pt I}}

\def\mapright#1{\smash{\mathop{\longrightarrow}\limits^{#1}}}

\def\Az{{\rm A\hskip -5.15pt\raise 1pt\hbox{$\scriptscriptstyle
\backslash$}} }
\def\nAz{{\nabla\hskip -6.6pt\raise 2.0pt
\hbox{$\scriptstyle\nabla$}} }
\def\az{{ \hbox{\nirm A}\hskip -4.9pt\raise 0.7pt\hbox{${\scriptscriptstyle
\backslash}$} }}

\def\lim{\mathop{\rm lim}}

\def\Aut{{\hbox{\rm Aut}}}
\def\End{{\hbox{\rm End}}}

\parindent=0pt
\font\eightrm                = cmr8
\font\eightsl                = cmsl8
\font\eightsy                = cmsy8
\font\eightit                = cmti8

\font\eighti                 = cmmi8
\font\eightbf                = cmbx8
\def\petit{\def\rm{\fam0\eightrm}
\textfont0=\eightrm %\scriptfont0=\sixrm \scriptscriptfont0=\fiverm
 \textfont1=\eighti %\scriptfont1=\sixi \scriptscriptfont1=\fivei
 \textfont2=\eightsy %\scriptfont2=\sixsy \scriptscriptfont2=\fivesy
 \def\it{\fam\itfam\eightit}
 \textfont\itfam=\eightit
 \def\sl{\fam\slfam\eightsl}
 \textfont\slfam=\eightsl
 \def\bf{\fam\bffam\eightbf}
 \textfont\bffam=\eightbf %\scriptfont\bffam=\sixbf
 %\scriptscriptfont\bffam=\fivebf
 \normalbaselineskip=9pt
 \setbox\strutbox=\hbox{\vrule height7pt depth2pt width0pt}
 \normalbaselines\rm}
\newdimen\refindent
\def\begref{\vskip0.5cm\bgroup\petit
\setbox0=\hbox{[Bi,Sc,So]o }\refindent=\wd0
\let\sl=\rm\let\INS=N}
\def\ref#1{\filbreak\if N\INS\let\INS=Y\vbox{\noindent\tbfontt
Refererences\vskip0.4cm}\fi\hangindent\refindent
\hangafter=1\noindent\hbox to\refindent{#1\hfil}\ignorespaces}
\long
\def\fussnote#1#2{{\baselineskip=9pt
\petit\noindent\footnote{\ \hskip -0.6cm  #1}{#2}}}
\parindent=0pt
%\rightline{hep-th/9605129}
%\footline={\hfil}
%\rightline{May 1996}
%\vskip 0.5cm
\phantom{mm}
\vskip 0.5cm
\leftline{\gross Dirac Operators and Clifford Geometry -}
\vskip 0.1cm
\leftline{\gross New Unifying Principles in Particle Physics ?}
\vskip 0.5cm
\leftline{\it Thomas Ackermann}
\vskip 0.3cm
\leftline{\eightrm Wasserwerkstra{\ss}e 37}
\leftline{\eightrm 68309 Mannheim, F.R.G.}
\leftline{\eightrm ackerm@euler.math.uni-mannheim.de}
\vskip 0.8cm
\leftline{\vbox{
%\hsize=5.0 true in
\petit\noindent
\bf Abstract.\rm\
In this lecture I will report on some recent
progress in understanding the relation of Dirac operators on Clifford
modules over an even-dimensional closed Riemannian manifold $M$\ and
(euclidean) Einstein-Yang-Mills-Higgs models.
}}
\vskip 0.6cm
{\sl Lecture at the ${\sl IV}^{th}$\ Conference on Clifford
Algebras and their
Applications in Mathematical Physics, Aachen, Germany, May 28 - 31,
1996}
\vskip 0.4cm

%{\mittel 1 Introduction}
%\vskip 0.3cm
Although being a gauge theory, it is well-known that the
classical theory of gravity as enunciated by Einstein
stands apart from the non-abelian gauge field theory of Yang, Mills and
Higgs,
which encompasses the three other fundamental forces: the
electromagnetic, weak and the strong interaction. General relativity
is governed by a variational principle associated with
the Lagrangian
$${S_{GR}={1\over 16\pi {\rm G}}\int_M\;*r_M}\eqno(1)$$
where G denotes Newton's constant, $*$\ is the Hodge star and
$r_M$\ is the scalar curvature of the space-time $M$, a closed
four-dimensional pseudo-Riemannian manifold of signature $(-, +\cdots +)$.
To describe a Yang-Mills-Higgs model the space-time manifold
$M$, apart from the metric, is endowed with additional structure (cf.
[IS]): Over
$M$\ we have both a principal bundle $P_G$\ with structure group a
compact Liegroup $G$\ and a Clifford module $\cal E$\ that
is assumed to furnish a representation $\rho\colon {\cal G}\rightarrow
\Aut_{C(M)}{\cal E}$\ of the group of
gauge transformations ${\cal G}$\ of $P_G$. Here $Aut_{C(M)}{\cal E}$\
are those automorphisms of ${\cal E}$\ which commute with the
Clifford action. Let $A$\ be a
connection on $P_G$\ with
curvature $R\in \Omega^2(M,\;ad(P_G))$\ \fussnote{${^1)}$}{Here $ad(P_G)$\ denotes
the vector bundle with fibre the Lie algebra $LG$\ associated
to $P_G$\ with respect to the adjoint representation
$Ad\colon G\rightarrow LG$.}, $\nabla^{\cal E}\colon
\Gamma({\cal E})\rightarrow \Gamma(T^*M\otimes {\cal E})$\ be the
associated
Clifford connection and $\varphi\in \Gamma(W)$\ a section of an
additional vector bundle $W$\ associated to $P_G$\ with induced connection
$\nabla^W$. Then a corresponding 
Yang-Mills-Higgs model
is based on the Lagrangian  
$${S_{YMH}=-{1\over g^2}\int_M tr\bigl(R\wedge *R\bigr) +\int_M
\Bigl(
\bigl(\nabla^W\varphi\wedge *\nabla^{W}\varphi\bigr)\!
+\!*V(\varphi)
\Bigr) +\int_M *(\psi,D_\varphi\psi)_{\cal E}}\eqno(2)$$\nobreak 
where 
$V\colon W\rightarrow \kz$\ is an invariant quartic polynomial,
$\psi\in \Gamma({\cal E})$\ and $D_\varphi\colon
\Gamma({\cal E})\rightarrow
\Gamma({\cal E})$\ de-\goodbreak
notes a Dirac-Yukawa operator associated to
$(\nabla^{\cal E},\;\varphi)$.
The constant $g$\ in (2) parametrizes the fibre metric $tr\colon
ad(P_G)\times ad(P_G)\rightarrow \kz$\ which is induced by the
Killing form on the Lie algebra $LG$\ of $G$\ and is called the
Yang-Mills coupling constant. If $G$\ is not simple, it is
possible to generalize (2) introducing a seperate coupling constant
for each simple factor.
\smallskip
With exception of the `pure'
Yang-Mills term  $S_{YM}=-{1\over g^2}
\int_M R\wedge *R$,
from a mathematical point of view the lagrangian
(2) looks highly artificial. So
we
briefly comment on its physical significance:
\smallskip
\parindent=10pt
\item{$\bullet$} The bosonic part of
a Yang-Mills-Higgs model which describes (nonabelian) gauge forces, is
defined by the first two terms of (2). The (covariant)
Klein-Gordon lagrangian $S_{KG}=\int_M \nabla^W\varphi \wedge *\nabla^W\varphi$\
and the Higgs potential
$S_\varphi=\int_M*V(\varphi)$\ for the Higgs field $\varphi\in \Gamma(W)$\
are added to the pure Yang-Mills part such that
the gauge bosons or connections $A^i\in \Omega^1(M,ad(P_G))$\
acquire masses. In the theory of electroweak interaction for example,
where we have $G_{\rm ew}=SU(2)\times U(1)$\ and $V(\varphi):=
{\lambda\over 4}
(\varphi,
\varphi)^2- {\mu^2\over 2} (\varphi,\varphi)$\ with $\varphi\in \Gamma(
M\times \kz^2)$, $(\;\cdot\; ,\;\cdot\;)$\ is the standard inner product
on $\kz^2$\ and $\lambda, \mu>0$.
%one thus obtains three massive - the
%$W^+,\;W^-$\ and $Z$\ - and one massless boson, the photon.   
\smallskip
\item{$\bullet$} The fermionic part $\int_M *(\psi,D_\varphi\psi)_{
\cal E}$\ which describes matter fields\ \fussnote{${^2)}$}{For example,
we have leptons
and quarks in the Standard Model.} is defined by the Dirac-Yukawa
operator $D_\varphi:=D+c_{\kl Y} \widetilde{\phi}(\varphi)$. Here
$D$\ is the Dirac
operator corresponding to the Clifford connection $\nabla^{\cal E}$\ and
$\widetilde{\phi}\colon W\rightarrow \End^-_{C(M)}{\cal E}$\ is a linear map.
The Yukawa coupling 
$c_Y\in \rz$\ gives rise to the fermion mass as soon as there exists
a non-vanishing $\varphi_0\in \Gamma(W)$\ which minimizes the
Higgs-potential $V$\ but is
only invariant under a subgroup ${\cal H }\subset {\cal G}$\ of
the group of gauge transformations corresponding to a subgroup
$H$\ of the gauge group $G$. For simplicity here we have assumed only one
fermion generation\fussnote{${^3)}$}{In general, with $N\in \nz$\ fermion
generations, the Clifford module $\cal E$\ corresponds to
${\cal E}\otimes \kz^N$. Thus, $c_Y\in \End(\kz^N)$\ mixing the different
fermion generations and we
obtain $\phi=\widetilde{\phi}(\varphi)\otimes c_Y$\ for
the Yukawa coupling. In the
Weinberg-Salam model of electroweak interaction
we have $N=3$\ corresponding to the three lepton families of
the electron $e$, muon $\mu$ and the tauon $\tau$.
%Further $M$\ is assumed
%to be spin, thus ${\cal E}=
%S{\hat\otimes}E\otimes \kz^N$\  and $\End_{C(M)}{\cal E}=\End E$\
%with $E=\kz_L^2\oplus \kz_R$\ being a trivial bundle corresponding
%to left and right-handed fermions.
}.  
\parindent=0pt
\smallskip The mass acquisition for the gauge bosons as well as for the
fermions of the theory by introducing the Higgs field $\varphi\in
\Gamma(W)$\ is called `spontaneous symmetry breaking'. For a mathematical
audience it might be considered as a physical counterpart of
reducing the $G$-principal bundle $P_G$, c.f. [NS]. It plays a central
r$\hat {\rm o}$le in the Weinberg-Salam model of electroweak interaction
where it produces three massive and one massless boson - the
$W^+,\;W^-,\;Z$\ and the photon - and gives also masses to the electron, the
muon and the tauon but not to the corresponding neutrinos.
\smallskip
As a particular Yang-Mills-Higgs model
based on the gauge group $G=SU(3)\times SU(2)\times U(1)$, nowadays
the Standard Model of
elementary particles is extraordinarily successful in describing
particle phenomenology. Nonetheless it suffers both from a technical and
concepual side: There are so many variables
(corresponding to coefficients in (2)) - eighteen in sum, among them
for example the gauge couplings, the masses of the bosons and
fermions etc. - which have to be experimentally determined
and put into the model. Conceptually, to mention only one
problem, it is not clear how to treat gravity in this con-\goodbreak
text. In principle
one should study an Einstein-Yang-Mills-Higgs model based on
the combined lagrangian
$${S=S_{\eightrm GR} \;+\; S_{\eightrm YMH}.}\eqno(3)$$
The corresponding Euler-Lagrange variational equations are coupled
equations for the gravitational and all bosonic and fermionic fields.
Because nowadays the best colliders in high-energy physics are only
able to observe scattering processes on a scale
determined by the masses of the heaviest fermions, for example by the
top-quark, and Newton's
constant G is so tiny compared to this scale, 
unfortunately gravitational effects among
individual particles are undetectable.
However there are other experimental data which seem to justify this
`Ansatz', cf. [D].  
\medskip
In this lecture I will report on some recent progress in understanding
the relationship of
Dirac operators and {\sl euclidean}
Einstein-Yang-Mills-Higgs models.
So we deal with a closed Riemannian manifold $M$\ of
even dimension $n$\ and
a Clifford module $\cal E$\ over $M$\ furnished with a
Clifford connection
$\nabla^{\cal E}\colon \Gamma(
{\cal E})\rightarrow \Gamma(T^*M\otimes {\cal E})$\ with
twisting curvature $R^{{\cal E}/S}\in \Omega^2(M,\End_{C(M)}{\cal E})$.
Tensoring $\cal E$\ with the 
$\zz_2$-graded trivial vector bundle $\kz^{1\vert 1}$\ whose
even and odd subbundles each
have complex rank one, obviously we obtain the Clifford module
${\bar {\cal E}}:={\cal E}\otimes \kz^{1\vert 1}$.
Furthermore denote by $\Psi {\rm DO}({\bar {\cal E}})$\ the space of
pseudo-differential operators on $\bar {\cal E}$. Then, one observes
the
\Theorem There exists a
Dirac operator $D_\phi\colon \Gamma({\bar{\cal E}})\rightarrow
\Gamma({\bar{\cal E}})$\ on ${\bar{\cal E}}:={\cal E}\otimes
\cz^{1\vert 1}$\ such that
$${res(D_\phi^{-n+2})\!\sim\!a_0\!
\int_M\! *r_M 
-a_1\!\int_M\! tr\bigl(R^{{\cal E}/S}\!\wedge\! *R^{{\cal E}/S}
\!\bigr)
+a_1\!\int_M\!
tr\bigl(\nabla^{\cal E}\phi\!\wedge\! *\nabla^{\cal E}\phi\!\bigr)
+ {a_1\over 2}\!\int_M\!*V(\phi)
}$$
where $res\colon \Psi {\rm DO}(\bar {\cal E})\rightarrow \kz$\ denotes
the non-commutative residue, $\phi\in \End_{C(M)}^-{\cal E}$, 
$V(\phi)=tr(\phi^4-\phi^2)$\ and the coefficients are $a_o={1
\over 12}$\ and $a_1={2\over {\rm rk}({\cal E}/S)}$, respectively,
where ${\rm rk}({\cal E}/S)$\ is the
rank of the (virtual) twisting part ${{\cal E}/S}$\ of the Clifford
module ${\cal E}$.
\smallskip\rm
Amazingly, this theorem not only offers an explanation of the
origin of both gravitational and Yang-Mills gauge symmetries, but
much more. Before arguing, I briefly scetch its proof (cf. [A3]):
\smallskip
Recall that the
non-commutative residue of
an operator $P\in \Psi {\rm DO}(E)$\ on a complex vector bundle $E$\
over
$M$\  
can be defined by
${res(P):={(2\pi)^{-n}}\int_{S^*M}
tr\bigl(\sigma^P_{-n}(x,\xi)\bigr) dx d\xi}$,
where $S^*M\subset T^*M$\ denotes the co-sphere bundle on $M$\ and
$\sigma^P_{-n}$\ is the component of order $-n$\ of the complete
symbol $\sigma^P:=\sum_i\;\sigma^P_i$\ of $P$, and is the only
non-trivial trace on the algebra of pseudo-differential operators
$\Psi {\rm DO}(E)$, cf. [W].
Given an elliptic operator $P$\ of order $d$\ and
$k\in \nz$\ with $n-k>0$, $res(P^{-{(n-k)\over d}})$\
can be related to the coefficient $h_k(P)$\
of $t^{k-n\over d}$\ in the asymptotic
expansion of $Tr\; e^{-tP}$. In particular, for a generalized laplacian
$\triangle$\ and $k=2$\ one obtains (cf. [KW], [A1])
$${res(\triangle^{-{n\over 2}+1})\;=\; c_n
\int_M\;*tr
\bigl({\hskip -0.6cm\hbox{\petit $1\over 6$}}r_M\eins_E - F\bigr)
}\eqno(4)$$\nobreak
with $c_n:= {\hskip -0.6cm\hbox{\petit
${(n-2)\over (4\pi)^{n\over 2}\cdot
\Gamma({n\over 2})}$} }$, since $h_2(\triangle)=\int_M\;*tr
\bigl({\hskip -0.6cm\hbox{\petit $1\over 6$}}r_M
\eins_E - F\bigr)$\ where $F\in \End\;E$\ orginates\hfil\goodbreak
from the decomposition $\triangle=\triangle^{{\hat\nabla}^E}+
F$, cf. [BGV]. Here
$\triangle^{{\hat\nabla}^E}$\ denotes
the connection laplacian corresponding to ${\hat\nabla}^E
\colon \Gamma(E)\rightarrow \Gamma(T^*M\otimes E)$.
\smallskip 
Now let $E={\cal E}$\ be a Clifford module and $D$\ a Dirac operator,
i.e. an odd-parity first order differential operator $D\colon
\Gamma({\cal E}^\pm)\rightarrow \Gamma({\cal E}^\mp)$\ such that
its square $D^2$\ is a generalized laplacian.
Using Quillen's theory of superconnections [Q]
on $\zz_2$-graded vectorbundles\ \fussnote{${^4)}$}{A superconnection
$\Az\colon \Omega^*(M, E)\rightarrow \Omega^*(M,E)$\ on a
$\zz_2$-graded vector bundle $E=E^ +\oplus E^-$\ can be defined as
a sum $\Az=\sum_{i\ge 0}\;\Az_{[i]}$\
of operators $\Az_{[i]}\colon \Omega^*(M,E)\rightarrow \Omega^{*+i}(M,E)$\
such that $\Az_{[1]}$\ is a connection on $E$\ which respects the
grading and $\Az_{[i]}\in \Omega^i(M,\End E)^-$\ for $i\ne 1$.
A superconnection $\Az\colon \Omega^*(M,{\cal E})^\pm
\rightarrow \Omega^*(M,{\cal E})^\mp$\ on a Clifford module
$\cal E$\ is called a Clifford superconnection, if it is compatible with
the Clifford action $c$, i.e.
${[\Az, c(a)]=c(\nabla a)}$\
for all $a\in \Gamma(C(M))$. For more details we recommend the
recent book [BGV].}
it is well-known that 
any Clifford superconnection
$\Az$\ on $\cal E={\cal E}^+ \oplus {\cal E}^-$\ uniquely determines a
Dirac operator ${D_{\Az}}$\
due to the following construction
$${D_{\az}\colon \Gamma({\cal E})\;\mapright{\az}\;
\Omega^*(M,{\cal E})\;
\mapright{\cong}\;\Gamma(C(M)\otimes
{\cal E})\;\mapright{c}\; \Gamma({\cal E}), }\eqno(5)$$
i.e. there is a one-to-one correspondence between Clifford
superconnections and
Dirac operators. The isomorphism is induced by
the quantisation map ${\bf c}\colon \Lambda^*(T^*M)\mapright{\cong} C(M)$\
and $c$\ denotes the given Clifford action of the Clifford
bundle $C(M)$\ on
${\cal E}$. Emphasizing this approach to Dirac operators one shows
the following generalization of Lichnerowicz's result (cf. [A2]):  
\Theorem Let $\Az=\Az_{[1]}+{\bar\Az}$\ be a Clifford
superconnection on a
Clifford module ${\cal E}$\ over an even-dimensional Riemannian
manifold $M$\ and let $D_\az:={\bf c}\circ \Az$\ denote the
corresponding Dirac operator. Then
$${D_\az^2\;=\;\triangle^{{\hat\nabla}^{\cal E}}\;+\;{r_M\over 4}\;+\;
{\bf c}\bigl(\Fz(\Az)^{{\cal E}/S}\bigr)\; +\;
P({\bar \Az})}\eqno(6)$$
where 
${\hat\nabla}^{\cal E}:=\Az_{[1]}+\beta({\bar\Az})$\   
determines the connection laplacian $\triangle^{{\hat
\nabla}^{\cal E}}$, $\Fz(\Az)^{{\cal E}/S}$\ denotes the
twisting supercurvature of $\Az$,
$P({\bar\Az}):= {\bf c}\bigl({\bar\Az}\bigr)^2 - {\bf c}\bigl(
{\bar\Az}^2\bigr) + ev_g\bigl(\beta({\bar\Az})\cdot
\beta({\bar\Az})\bigr)$\ and
$\beta({\bar\Az})\in \Omega^1(M,\End\;
{\cal E})$\ is locally defined by $\beta({\bar\Az}):= dx^k\otimes
{\bf c}\bigl(i(\partial_k){\bar\Az}\bigr)$.
\smallskip\rm
Here the dot `$\cdot$' indicates the fibrewise defined product in the
algebra bundle $T(M)\otimes \End\;{\cal E}$\ with $T(M)$\ being the
tensor bundle of $T^*M$. The endomorphism $P({\bar\Az})\in
\Gamma(\End {\cal E})$\ depends only on the higher degree parts
$\Az_{[i]},\ i\ge 2$\ of the Clifford superconnection. For example,
calculation of $P({\bar\Az})$\ for
$\Az:=\Az_{[0]}+\Az_{[1]}+\Az_{[2]}$\ where the
two-form part is given by
$\Az_{[2]}:={1\over 2}\;dx^i\wedge dx^j\otimes \omega_{ij}$\
with $\omega_{ij}\in \End_{C(M)}{\cal E}$\ for all $i,\;j$\
yields
$${P({\bar\Az})=-2g^{ij}\;{\bf c}(dx^k\wedge dx^l)\;\omega_{ik}\omega_{jl}
- g^{ij} g^{kl}\; \omega_{ik}\omega_{jl}.}\eqno(7)$$
According to (4) it is evident that the generalized Lichnerowicz formula
(6) is the main tool to compute the non-commutative residue
$res(D^{-n+2}_\az)$\ of a Dirac operator $D_\az$. Since it can be shown
that 
$tr\bigl({\bf c}(\Fz(\Az)^{{\cal E}/S})\bigr)=tr(\Az_{[0]}^2)$\
(cf. [A3]) these results imply 
$${res(D_\az^{-n+2})\;=\; -c_n\cdot
\int_M\;*\;tr
\bigl({\hskip -0.6cm\hbox{\petit $1\over 12$}}r_M\eins_{\cal E} +
\Az^2_{[0]}+ P({\bar\Az})\bigr).
}\eqno(8)$$\nobreak
Notice that $res(D_{f^*\az}^{-n+2})=res(D_{\az}^{-n+2})$\ holds for all
$f\in \Aut_{C(M)}{\cal E}$\ which can be seen as the `source' of
gauge invariance. Computing (8) for our example
$\Az:=\Az_{[0]}+\Az_{[1]}+\Az_{[2]}$\ mentioned above
where $P({\bar\Az})$\ is given by equation (7) we obtain
$${res(D_\az^{-n+2})\;=\; -c_n\cdot
\int_M\;\Bigl({\hskip -0.6cm\hbox{\petit $rk({\cal E})\over 12$}} *r_M +
tr\bigl(\Az_{[0]}\wedge *\Az_{[0]}\bigr)+2tr\bigl(\Az_{[2]}\wedge *
\Az_{[2]}\bigr)\Bigr).}\eqno(9)$$
As usual, $tr(\;\cdot \wedge\! *\;\cdot\;)$\ denotes the bilinear form
defined by the standard inner product $(\;\cdot\wedge\! *\;\cdot\;)$\
on the exterior bundle $\Lambda^*T^*M$\
combined with the complex trace
$tr$\ on the endomorphism bundle $\End\; {\cal E}$. Obviously
$tr(\;\cdot\wedge\! *\;\cdot\;)$\ is symmetric but in general
neither real nor definit. 
\smallskip
Finally I turn back to
particular geometric situation of theorem 1: Consider
the odd endomorphism $J\in \End^-(\cz^{1
\vert 1})$\
which is defined  by $J(z_1,z_2)=(z_2,-z_1)$\ for all $(z_1,z_2)\in C^{1\vert 1}\cong
\cz\oplus \cz$. Then $J^2=-\eins_{\kz^{1\vert 1}}$\ holds and
obviously we have an extension ${\cal J}:=
\eins_{\cal E}{\hat \otimes} J$\ to
the tensor bundle ${\bar {\cal E}}={\cal E}\otimes \cz^{1\vert 1}$\
with the same properties.
There are natural `even' and `odd' inclusions $\iota,j\colon
\End {\cal E}
\hookrightarrow \End({\cal E}\otimes \cz^{1\vert 1})$\ defined by
$\iota(F):= F{\hat\otimes}\eins_{\kz^{1\vert 1}}$\ and
$j(F):=F{\hat\otimes}J$, respectively. Now assume that 
$\nabla^{\cal E}$\ is a Clifford connection on ${\cal E}$\ and let
$\nabla^{\bar{\cal E}}:=\bigl({\nabla^{\cal E}\atop 0}{0\atop
\nabla^{\cal E}}\bigr)$\ be the induced Clifford connection on
${\bar{\cal E}}$. To any odd endomorphism $\phi\in \Gamma(\End_{C(M)}^-
{\cal E})$\ we can define
an even $\End_{C(M)} {\bar{\cal E}}$-vallued one form
$\alpha:=\sum_\mu dx^\mu\otimes j(\phi)$. Hence, $\widetilde{\nabla}:=
\nabla^{\bar{\cal E}}+\alpha$\ is a Clifford connection 
and $\widetilde{\Az}:=\iota(\phi)+\widetilde{\nabla}$\ is a Clifford
superconnection on $\bar {\cal E}$\ with twisting supercurvature
$\Fz(\widetilde{\Az})^{{\bar{\cal E}}/S}=\bigl(\widetilde{\nabla}^2
\bigr)^{{\bar {\cal E}}/S} +
\widetilde{\nabla}\iota(\phi) +\iota(\phi^2)$. We define
$${\Az\;:=\;\iota(i\phi)\;+\;\nabla^{\bar{\cal E}}\;+\;
{\cal J}\Fz(\widetilde{\Az})^{{\bar {\cal E}}/S}.}\eqno(10)$$
Since $\Az$\ is a Clifford superconnection on $\bar {\cal E}$,
the corresponding Dirac operator $D_\phi:=D_\az$\ is shown to
verify 
theorem 1 by using (9),
the decomposition $\Az=\Az_{[0]} + \Az_{[1]} + \Az_{[2]}$\
with    
$$\matrix{\Az_{[0]} =\bigl(\iota(i\phi)+{\cal J}\iota(\phi^2)
\bigr), &\Az_{[1]} =\bigl(\nabla^{\bar{\cal E}} +
{\cal J}\widetilde{\nabla}\iota(\phi)
\bigr), &\Az_{[2]} =\bigl({\cal J}\iota(R^{{\cal E}/S}) +
{\cal J} d^{\nabla^{\bar{\cal E}}}\alpha\bigr), \cr}\eqno(11)$$
and the properties of ${\cal J}$. For more details I refer once
more to [A3].
\medskip
Let me now illustrate the power of theorem 1 when
applied to particle physics
by means of a `toy model' which is strongly related to
the celebrated Weinberg-Salam model of electroweak interaction: 
\smallskip
We choose a four-dimensional closed Riemannian manifold $M$,
for convinience endowed with a spin structure $S$\ and let $P_{G_{ew}}:=
M\times G_{\rm ew}$\ be a trivial principal bundle with structure group
$G_{\rm ew}:=SU(2)\times U(1)$. To describe fermionic
fields $\psi\in \Gamma({\cal E})$\ where
${\cal E}:=S\otimes E$, let $E=P_{G_{\rm ew}}\times_\rho \cz^2_L
\oplus \cz_R$\ be the bundle
associated to $P_{G_{\rm ew}}$\ with respect to
the representation $\rho(U,\;e^{i\theta})(v\oplus w):=Ue^{iy_L\theta}v
\oplus e^{iy_R\theta}$\ for all $(U,e^{i\theta})\in G_{\rm ew}$\ and
$(v\oplus w)\in \cz_L^2\oplus \cz_R$. Here $\cz^2_L,\; \cz_R$\ correspond
to left- and right-handed fermions with hypercharges $y_L,\;y_R\in \zz$,
respectively\ \fussnote{${^5)}$}{Suggestively we write $\psi_L=\bigl({\nu_e
\atop e_L}\bigl)\in \Gamma(S\otimes \cz_L^2),
\;\psi_R=e_R\in \Gamma(S\otimes \cz)$. To make things easier, notice that
$N=1$, i.e. we assume
to have only one fermion generation.}. Furthermore the Higgs-field
$\varphi=\bigl({\varphi_1\atop \varphi_2}\bigr)\in \Gamma(W)$\ - with
$W:=M\times \cz^2$\ being associated to $P_{G_{\rm ew}}$\ with respect
to the representation with hypercharge $y_\varphi\in \zz$\ - is
incorporated into the model by the linear map
$$\matrix{\widetilde{\phi}\colon \End(M\times \cz^2)\longrightarrow
\End^{-}_{C(M)}{\cal E},&\quad &\varphi\;\mapsto\;\widetilde{\phi}(
\varphi):=\hskip -0.4cm
\hbox{\petit $\left( \matrix{ 0 &0 &\varphi_1\cr
                              0 &0 &\varphi_2\cr
                             {\bar\varphi}_1 &{\bar\varphi}_2 &0\cr}\right)
$}\cr}\eqno(12)$$  
which is determined by the Yukawa coupling $\phi(\varphi)=c_e\widetilde{
\phi}(\varphi)$, i.e. the equality $(\psi,\phi(\varphi)\psi)=(\psi,c_e
\widetilde{\phi}(\varphi)\psi)$\ holds. Here the coupling constant $c_e
\in \rz$\
gives rise to the mass of the electron $m_e$. Note that until now
we have not fixed neither the hypercharges $y_L,\; y_R$\ for the fermions
nor $y_\varphi$\ for the Higgs. Let $\nabla^W\colon \Gamma(W)\rightarrow
\Gamma(T^*M\otimes W)$\ be a connection and $\nabla^{\cal E}\colon
\Gamma({\cal E})\rightarrow \Gamma({\cal E})$\ be a Clifford connection,
both induced by a principal connection 
$A$\ on $P_{G_{\rm ew}}$. Interestingly one obtains (cf. [A4]):    
\Lemma Let $\nabla^{\cal E}\colon
\Gamma(\End {\cal E})\rightarrow \Gamma(T^*M\otimes\End {\cal E})$\
also denote the connection on the endomorphism bundle $\End {\cal E}$.
Then $tr\bigl(\nabla^{\cal E}\widetilde{\phi}(\varphi)\wedge * \nabla^{
\cal E}\widetilde{\phi}(\varphi)\bigr)=4(\nabla^W\varphi\wedge *\nabla^W
\varphi)$\ holds iff $y_\varphi=-y_L$\ and $y_\varphi=y_L-y_R$.
\smallskip\rm 
Let us now consider a
variant of the Dirac operator $D=D_\az$\ on ${\bar {\cal E}}={\cal E}
\otimes \cz^{1\vert 1}$\ defined by the
Clifford superconnection (10), more precisely the
family $D_{(c_e, b_w, b_y)}$\ of Dirac operators  
corresponding to the Clifford superconnections
$\Az_{(c_e,b_w,b_y)}:=\iota(i \phi(\varphi))+\nabla^{\bar{\cal E}}+
{\cal J}\iota(B)\Fz(\widetilde{\Az})^{{\bar {\cal E}}/S}$\
uniquely determined by
$$\matrix{\phi(\varphi)\!:=\!c_e\widetilde{\phi}(\varphi)\!\in\!
\End_{C(M)}^-{
\cal E}, &B\!:=\!\hskip -0.6cm
\hbox{\petit $\left(\matrix{ b_w &0 &0 \cr
                              0 &b_w &0\cr
                              0 &0  &b_y\cr}\right)$}\!\in\!
\End_{C(M)}^+{\cal E},
&(c_e,b_w,b_y)\in \rz^3.\cr}\eqno(13)$$
Since it can be shown that $tr\bigl(\iota(i\phi(\varphi))+{\cal J}\iota(
B\phi(\varphi)^2)\bigl)=2c_e^4b_w b_y(\varphi,\varphi)^2 - 2c_e^2 (\varphi,
\varphi)$, one now uses lemma 3 and theorem 1 to obtain the 
\Corollary Let ${\bar {\cal E}}:={\cal E}\otimes \cz^{1\vert 1}$\
be the particular Clifford module as defined above and $D_{(c_e,b_w,b_y)}
\colon \Gamma({\bar {\cal E}})\rightarrow \Gamma({\bar {\cal E}})$\ be the
family of Dirac operators corresponding to (13). Then
$${res(D^{-2}_{(c_e,b_w,b_y)})\sim\cases{ 
 l^{-2}_p\int_M *r_M
- A_w\int_M *tr(G_{\mu\nu}G^{\mu\nu})
&\ \cr
\noalign{\vskip 0.2cm}
\ \; -A_y \int_M *tr(F_{\mu\nu}F^{
\mu\nu})+ L \int_M *(\nabla^W_\mu\varphi,\nabla^W_\mu\varphi)
&\ \cr
\noalign{\vskip 0.2cm}
\ \ \;
+ K_1 \int_M *(\varphi,\varphi)^2 - K_2 \int_M *(\varphi,\varphi)
&\ \cr}}\eqno(14)$$
where $G$\ and $F$\ denote the $L(SU(2))$- and
$L(U(1))$- components of the twisting curvature $R^{{\cal E}/S}\in
\Omega^2(M,L(G_{\rm ew}))$\ corresponding to the Clifford connection
$\nabla^{\cal E}$\ under the decomposition of the Lie algebra
$L(G_{\rm ew})=L(SU(2))\oplus L(U(1))$\ and
with coefficients $A_w:=\alpha_1 b_w^2,\;
A_y:=\alpha_1 b_y^2,\; L:=4\alpha_1 b_w b_y c_e^2,\;K_1:=\alpha_1 b_w
b_y c_e^4,\;K_2:=\alpha_1 c_e^2$\ with $\alpha_1\sim a_1$.
\smallskip\rm
Notice that we have introduced a physical lenght scale in units
fixed by the Planck lenght $l_p:=\sqrt{16\pi {\rm G}}$. Thus, (14) can be
interpreted as the bosonic parts corresponding to (euclidean)
Einstein-Yang-Mills-
Higgs models (3) with one fermion generation which are classified
by $(\alpha_1, b_w, b_y, c_e)\in \rz^4$.
\smallskip 
One immediate
consequence of this corollary occurs if one `switches off' the
gravitational force, i.e. we are in flat space, and fixes
the hypercharge of the left-handed fermions by $y_L=-1$. Hence
lemma 3 implies $y_\varphi=1$\ and $y_R=-2$. So we are exactly in the
case of the Weinberg-Salam model with one lepton generation
${(\nu_e, e_L, e_R)}$. With exception of the `geometrical input',
which, among others, concerns also the above mentioned combination of
the hypercharges $(y_L, y_R, y_\varphi)$, it is well-known that
this model can be parametrized by
$${\bigl(g_w,\; g_y,\; c_e,\;\lambda,\;\mu\bigr)\in \rz^5,}\eqno(15)$$
cf. [N]. Here $g_w,\; g_y$\ are the coupling constants of the
$SU(2)$\ and $U(1)$\ gauge fields, respectively, $c_e$\ is the
Yukawa-coupling and $\lambda,\;\mu$\ are the constants which enter into
the Yang-Mills-Higgs lagrangian (2) via the Higgs potential $V(
\varphi)$\ as already mentioned. However, identification of our
coefficients $(A_w, A_y, L, K_1, K_2)$\ to the standard ones
$$\matrix{\widetilde{A}_w:={1\over 2g_w^2}, &\widetilde{A}_y:={1\over
4g_y^2}
&\widetilde{L}:=1, &\widetilde{K}_1:={\lambda\over 4},
&\widetilde{K}_2:={\mu^2\over 2} \cr}\eqno(16)$$
of the
bosonic part of (2)\ \fussnote{${^6)}$}{Note that one has to take into
account that $U(1)$\ couplings are conventionally normalized differently
than $SU(n)$\ gauge couplings.} yields
$${\lambda\;=\;{1\over 2\sqrt{2}}\;g_w\cdot g_y.}\eqno{(17)}$$   
Thus, by our approach (14) we are indeed able to reduce the variables
(15) of the model. In its turn $\lambda$\ enters
into the relation
${m_\varphi^2\over m_w^2}= {8\over g_w^2}\cdot \lambda$\ of the Higgs
and the $W$-boson masses $m_\varphi,\; m_w$. Consequently we
have $m_\varphi^2=2\sqrt{2}\cdot
{g_y\over g_w}\cdot m_w^2=2\sqrt{2}\cdot {\rm tan}\theta_w\cdot
m_w^2$\ where $\theta_w$\ denotes the so-called Weinberg angle, cf.
[N]. Since both ${\rm sin}^2\theta_w= 0.2325\pm 0.0008$\ and
$m_w= 80.22\;(26)$\ GeV are determined by scattering experiments,
our simplified model provides us with the concrete (but unrealistic)
prediction
$${m_\varphi\;\approx\; 100.1\ {\rm GeV}.}\eqno(18)$$
\medskip   
So what have we learned until now ? Given a `gauged'
Clifford module ${\cal E}$\ - i.e. a Clifford module which furnishes
a representation $\rho\colon {\cal G}\rightarrow \Aut_{C(M)}{\cal E}$\
of the group of gauge transformations $\cal G$\ of a given principal
bundle $P_G$\ with structure group $G$\ - and a
Dirac-Yukawa operator $D_\varphi\colon\Gamma({\cal E})\rightarrow
\Gamma({\cal E})$,
one should take $res(D_\phi^{-2})$\ (where
$D_\phi$\ denotes the particular Dirac operator
on ${\bar {\cal E}}:={\cal E}\otimes \cz^{1\vert 1}$\
determined by (10) and (13), respectively) as a {\sl definition}
of the bosonic action $S_{\rm b}$\ of an Einstein-Yang-Mills-Higgs
model. Moreover it is easily verified that in terms of this
Dirac operator $D_\phi$\ the fermionic
action $S_{\rm f}$\ can be defined by $S_{\rm f}:=\int_M*(\Psi,D_\phi
\Psi)_{\bar {\cal E}}$\ with $\Psi:={1\over \sqrt{2}}\sum_i\psi\otimes
e_i \in \Gamma({\bar{\cal E})
}$, $\{e_1,\;e_2\}$\ denotes a orthonormal base of $\cz^{1\vert 1}$.
Thus, we suggest to replace (3) by the {\sl definition}
$${S:=res(D_\phi^{-2})\;+\;\int_M*(\Psi,D_\phi\Psi)_{\bar {\cal E}},}\eqno(
19)$$ 
cf. [A3]. Writen in this most advanced form, the particular Dirac
operator $D_\phi$\ on ${\bar {\cal E}}$\ can be
identified as the origin of both gravitational and
Yang-Mills gauge symmetries. At least in the case of our `toy model'
considered above, there is little extra information captured by this
approach (19): On the geometrical side, calculation of the kinetic
term for the Higgs
involves constrains on the representations of the $U(1)$-part of the
gauge group $G_{ew}$\ which result in the correct relations
for the hypercharges
$y_L,y_R$\ and $y_\varphi$. 
Moreover, the free variables of the model can be reduced which, in
addition, constrains the Higgs mass $m_\varphi$\ to be $100.1$\ GeV,
approximately. For the case of the full Standard model we refer to the
forthcoming work [A4].
\vskip 0.5cm
\centerline{\TT Conclusion}
\bigskip
In this lecture I tried to show that
Dirac operators on certain Clifford modules indeed could provide new
insights into the origin of gauge symmetries in particle physics.
Our approach
(19) might also fit to `unify', or at least connect, some
different efforts in this field: Emphasizing
the one-to-one correspondence of Dirac operators
and Clifford superconnections there are relations
to the Ne'eman-Sternberg approach to
the Standard model [NS], for example the geometrical interpretation
of the Higgs as a component of a superconnection.   
Furthermore, as our approach (19) to gravity coupled with
matter relies completly on terms of differential geometry which
have an analogue
in the noncommutative world,
(19) can be seen as a `starting point' to define Einstein-Yang-Mills-
Higgs models on noncommutative quantum spaces [C]. Note that the main
idea concerning (19) can already be found in [AT1], [AT2].
Unfortunately in [AT2]
the Dirac operator $\widetilde{D}$\ is incorrect. However this will be
treated elsewhere.
\begref
\ref{[A1]} T.Ackermann, \sl A Note on the Wodzicki Residue
\rm , to appear in Journ. of Geom. and Physics
\ref{[A2]} T.Ackermann, \sl Supersymmetry and the generalized
Lichnerowicz formula,  dg-ga/9601004
\ref{[A3]} T.Ackermann, \sl Dirac Operators and Particle Models,
Part I: Yang-Mills-Higgs coupled to Gravity, 
to appear 
\ref{[A4]} T.Ackermann, \sl Dirac Operators and Particle Models,
Part II, forthcoming 
\ref{[AT1]} T.Ackermann, J.Tolksdorf, \sl The generalized
Lichnerowicz formula and analysis of Dirac operators, Journ. reine
angew. Math. {\bf 471} (1996), 23-42 
\ref{[AT2]} T.Ackermann, J.Tolksdorf, \sl A Unification of
Gravity and Yang-Mills-Higgs Gauge Theories, CPT-95/P.3180
\ref{[BGV]} N.Berline, E.Getzler, M.Vergne, \sl Heat kernels and
Dirac operators\rm , Springer (1992)
\ref{[C]} A.Connes, \sl Gravity coupled with matter and the
foundation of non commutative geometry, hep-th/9603053
\ref{[D]} T.Damour, \sl General Relativity and Experiment, Proc. of
the XI$^{th}$\ International Congress of Math. Physics, Editor D.
Iagolnitzer, Intern. Press (1995)
\ref{[IS]} B.Iochum, T.Sch{\"u}cker, Yang-Mills-Higgs versus
Connes-Lott, Com. Math. Phys. {\bf 178} (1996), 1-26
\ref{[N]} O.Nachtmann, Elementarteilchenphysik, Ph{\"a}nomene
und Konzepte, Vieweg (1986)
\ref{[NS]} Y.Ne'eman, S.Sternberg, Internal Supersymmetry
and Superconnections, Sympl. Geom. and Math. Physics, Birkh{\"a}user
(1991)\nobreak
\ref{[Q]} D.Quillen, \sl Superconnections and the Chern Character\rm ,
Toplology {\bf 24} (1985), 89-95\nobreak
\ref{[W]} M.Wodzicki, \sl Non-commutative residue I\rm , LNM {\bf 1289}
(1987), 320-399\goodbreak
\bye
\ref{[Wa]} M.Walze, \sl Nicht-kommutative Geometrie und

\bye
\vskip 0.5cm
{\mittel 2 Some machinery ...}
\ukneu
\bye
\vskip 0.5cm
{\mittel 3 ... and the theorem}
\ukneu
\vskip 0.5cm
{\mittel 4 The Standard Model. And beyond ?}
\bye